\documentclass[aip,cha,amsmath,amssymb,reprint,author-year,author-numerical]{revtex4-1}
\usepackage{epsfig}
\usepackage{color}
\usepackage{graphicx}
\usepackage{dcolumn}
\usepackage{bm}
\input epsf

\usepackage[normalem]{ulem} 

\graphicspath{{figures/}}

\begin{document}


\title{Uncovering low dimensional macroscopic chaotic dynamics of large finite size complex systems}

\author{Per Sebastian Skardal}
\email{persebastian.skardal@trincoll.edu}
\affiliation{Department of Mathematics, Trinity College, Hartford, CT 06106, USA}

\author{Juan G. Restrepo}
\affiliation{Department of Applied Mathematics, University of Colorado, Boulder, CO 80309, USA}

\author{Edward Ott}
\affiliation{Department of Physics, University of Maryland, College Park, MD 20742, USA}

\begin{abstract}
In the last decade it has been shown that a large class of phase oscillator models admit low dimensional descriptions for the macroscopic system dynamics in the limit of an infinite number $N$ of oscillators. The question of whether the macroscopic dynamics of other similar systems also have a low dimensional description in the infinite $N$ limit has, however, remained elusive. In this paper we show how techniques originally designed to analyze noisy experimental chaotic time series can be used to identify effective low dimensional macroscopic descriptions from simulations with a finite number of elements. We illustrate and verify the effectiveness of our approach by applying it to the dynamics of an ensemble of globally coupled Landau-Stuart oscillators for which we demonstrate low dimensional macroscopic chaotic behavior with an effective 4-dimensional description. By using this description we show that one can calculate dynamical invariants such as Lyapunov exponents and attractor dimensions. One could also use the reconstruction to generate short-term predictions of the macroscopic dynamics.
\end{abstract}


\maketitle

\begin{quotation}
The search for emergent low-dimensional behavior in the macroscopic dynamics of large systems of dynamical units is an important issue in many scientific and technological contexts. Explicit analytical low dimensional macroscopic descriptions have been found for certain classes of phase-oscillator systems with sinusoidal coupling~\cite{Watanabe1993PRL,Ott2008Chaos}. However, such analytically derived descriptions remain elusive for more general cases, and it is probably unrealistic to expect that similar analytical methods will be found for general situations. Here we demonstrate a method for identifying low dimensional behavior in macroscopic dynamics of large coupled systems exhibiting complicated dynamics using techniques originally designed for denoising chaotic time series~\cite{Kantz2003,Sprott2003}. As an illustration, we study in detail a system of globally-coupled Landau-Stuart oscillators~\cite{Pikovsky2003} in the chaotic regime for which we reconstruct the low-dimensional dynamics. The basic idea is to remove finite-size fluctuations so as to reveal the underlying low dimensional macroscopic dynamics, thus allowing for accurate computation of dynamical invariants~\cite{Ott2002}.
\end{quotation}

\section{Introduction}\label{sec1}

Large systems of coupled dynamical units serve as important models for phenomena in a wide range of disciplines including physics, engineering, biology, and chemistry~\cite{Kuramoto}. In a suitable thermodynamic limit (i.e., where the number $N$ of interacting units goes to infinity in an appropriate limiting process) low dimensional descriptions have been {\it explicitly} found for certain families of sinusoidally-coupled phase oscillator systems, first in the case of identical oscillators~\cite{Watanabe1993PRL,Watanabe1994PhysD,Goebel1995PhysD,MarvelChaos2009}, then for heterogeneous oscillators~\cite{Ott2008Chaos,Ott2009Chaos}. These analytical breakthroughs made possible the analysis of various generalizations of the Kuramoto model of phase oscillators, including the behavior of systems with time-delays~\cite{Lee2009PRL}, external forcing~\cite{Childs2008Chaos}, communities~\cite{Pikovsky2008PRL,Skardal2012PRE}, more general network structures~\cite{Restrepo2014EPL,Skardal2015PRE}, and pulse-coupled units~\cite{Pazo2014PRX,Luke2014FCN,Laing2014PRE,Montbrio2015PRX,Laing2015SIAM}. To date, no such dimensionality reduction is known for the case where more general kinds of dynamical units are coupled. In general, at finite $N$, guided by the results from Refs.~\cite{Ott2008Chaos,Ott2009Chaos,Lee2009PRL,Childs2008Chaos,Pikovsky2008PRL,Skardal2012PRE,Restrepo2014EPL,Skardal2015PRE,Pazo2014PRX,Luke2014FCN,Laing2014PRE,Montbrio2015PRX,Laing2015SIAM}, one might expect that, at large, but not too large $N$, the behavior would often (not always) resemble an effective superposition of high dimensional and low amplitude noise-like dynamics, superposed upon low dimensional macroscopic behavior. In many cases, the noise-like component can obscure the identification of the underlying low dimensional dynamics. Moreover, the lack of a low dimensional description of the macroscopic dynamics combined with the presence of finite-size fluctuations often prevents the computation of important dynamical invariants for the macroscopic behavior. 

\begin{figure*}[t]
\centering
\epsfig{file =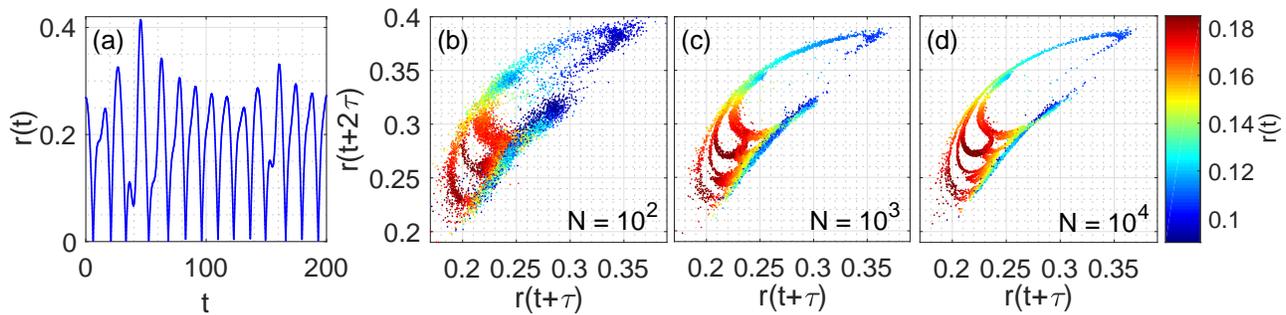, clip =,width=0.95\linewidth }
\caption{(Color online) (a) Chaotic time series of the order parameter $r(t)$ for the system of size $N=10^3$ described in Eq.~(\ref{eq:01}) with coupling strength $K=0.885$ and natural frequencies uniformly distributed in $[-\pi\Delta/2,\pi\Delta/2]$ for $\Delta = 0.64$. (b)--(d) Surface of sections constructed from the $d_T=4$ dimensional time delay embedding with time delay $\tau = 2.8037$ where the points in panels (b)--(d) are plotted at downward-moving ($dr(t+3\tau)/dt<0$) piercings of the hyperplane $r(t+3\tau)=0.2$ for systems of size $N=10^2$, $10^3$, and $10^4$.}\label{fig1}
\end{figure*}

In this paper we show that low dimensional dynamics in large systems of coupled oscillators can be identified by reconstructing these dynamics using techniques first developed for denoising experimental chaotic time series. In general, we assume the existence of an observable time series from which an appropriate time-delay embedding~\cite{Takens} can be generated which in turn can be used to build a discrete mapping via an appropriate surface of section~\cite{Ott2002}. As discussed above, this mapping will be effectively noisy due to finite-size fluctuations. We then reconstruct the low-dimensional dynamics using a denoising technique that makes local estimations for the evolution of the state variables by averaging out noise in the original dataset. The reconstructed low dimensional dynamics can then be used to accurately calculate important dynamical invariants such as fractal dimensions and Lyapunov exponents. As an illustrative example, we apply this procedure to a system of globally coupled Landau-Stuart oscillators that we find effectively exhibits macroscopic chaos with substantial superposed noise-like finite-size fluctuations.

The remainder of this paper is organized as follows. In Sec.~\ref{sec2} we describe the system of Landau-Stuart oscillators and summarize its dynamics and observed finite-size fluctuations. In Sec.~\ref{sec3} we describe our method for reconstructing the low dimensional dynamics. In Sec.~\ref{sec4} we demonstrate our method's utility in calculating invariants of the macroscopic dynamics of the Landau-Stuart system. 
In Sec.~\ref{sec5} we conclude with a discussion of our results.

\section{System Dynamics}\label{sec2}

In this paper we use as our primary example a system of $N$ globally-coupled Landau-Stuart oscillators whose dynamics are governed by 
\begin{align}
\dot{z}_n&=z_n(1-|z_n|^2+i\omega_n)+\frac{K}{N}\sum_{m=1}^N(z_m-z_n)\nonumber\\
&=z_n(1-|z_n|^2+i\omega_n)+K(\overline{z}-z_n).\label{eq:01}
\end{align}
The complex variable $z_n$ describes the (complex) state of oscillator $n$, $\omega_n$ is the natural frequency of oscillator $n$, $K$ is the global coupling strength, and $\overline{z}=(1/N)\sum_{m=1}^N z_m$ is the global mean-field. Importantly, a rich set of dynamical phenomena have been observed and studied in Eq.~(\ref{eq:01}) and its variations, including what appears to be macroscopic chaos~\cite{Matthews1990PRL,Mirollo1990JSP,Matthews1991PhysD,Hakim1992PRA,Nakagawa1993PTP,Takeuchi2013JPA,Ku2015Chaos}. For the purpose of this paper we restrict our attention to the dynamics that emerge with a coupling strength of $K=0.885$ and natural frequencies that are uniformly distributed (and evenly spaced) in $[-\pi\Delta/2,\pi\Delta/2]$ for $\Delta = 0.64$. These choices result in behavior that is suggestive of chaotic dynamics, as can be seen in the time series of the order parameter $r(t)=|\overline{z}(t)|$ plotted in Fig.~\ref{fig1}(a) for a system of size $N=10^3$. 

To analyze this chaotic state, we proceed by finding a suitable time-delay embedding for the variable $r(t)$. As discussed in the next section, the minimum embedding dimension for the denoised low dimensional dynamics can be found using several methods~\cite{Ding1993PhysD,Cao1997PhysD}; here we use the method of false nearest neighbors~\cite{Kennel1992PRA,Abarbanel1993PRE} resulting in a minimum embedding dimension of $d_{T}=4$. Moreover, we find that $\tau=2.8037$ is a suitable time delay. This embedding then yields a $d_T=4$ dimensional state variable $\bm{y}(t)=[r(t),r(t+\tau),r(t+2\tau),r(t+3\tau)]^T$. We next construct a discrete mapping using a surface of section collected at the downward piercings of the hyperplane $r(t+3\tau)=0.2$ resulting in a sequence of $d_T-1=3$ dimensional state variables $\bm{x}_n=[r(t_n),r(t_n+\tau),r(t_n+2\tau)]^T$ where $t_n$ is the time of the $n^{\text{th}}$ piercing of the surface. In Figs.~\ref{fig1}(b)--(d) we plot attractors found using this surface of section for system sizes $N=10^2$, $10^3$, and $10^4$, plotting $r(t+\tau)$ and $r(t+2\tau)$ on the horizontal and vertical axes, respectively, and denoting the value of $r(t)$ by color. Comparing the three cases, we observe a strong dependence of the noise intensity on the system size. In the $N=10^2$ case the noise level is particularly strong, corrupting virtually all the detail in the attractor. As the system size increases to $N=10^3$ and $10^4$ more detail can be seen. However, it is clear that substantial noise persists even for $N=10^4$. We note that while the embedding dimension $d_T=4$ provides an upper bound for the dimensionality of the macroscopic dynamics of the system, this embedding dimension is often larger than the actual dimension of the system~\cite{Takens}. Moreover, the complex behavior and degree of finite-size fluctuations impede any direct quantification of other properties of the dynamics, e.g., dimension of the attractor and stretching and contracting of trajectories. 

\begin{figure}[t]
\centering
\epsfig{file =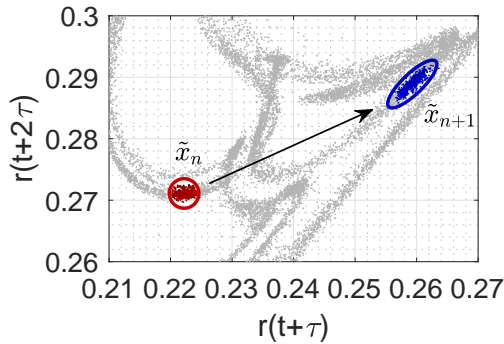, clip =,width=0.95\linewidth }
\caption{(Color online) Illustration of the noisy forward mappings $\tilde{\bm{x}}_n\mapsto\tilde{\bm{x}}_{n+1}$ used in the least squares calculation of $\bm{F}_{\text{loc}}$ from the example presented in Fig.~\ref{fig1}.}\label{fig2}
\end{figure}

\section{Low Dimensional Reconstruction}\label{sec3}
To better quantify the macroscopic system properties from cases like that illustrated above, we seek to reconstruct the low dimensional dynamics directly from a dataset. The method presented here is a variation on existing methods~\cite{Kantz2003} developed for denoising experimental chaotic time series, and thus reconstructs a denoised low-dimensional representation of the dynamics. We describe this method in general, assuming the existence of an observable noisy time series $r(t)$ for which a suitable time-delay embedding with embedding dimension $d_T$ can be found. Moreover, we assume that a mapping can be constructed using a general surface of section resulting in the $(d_T-1)$-dimensional state vector $\tilde{\bm{x}}$, whose components we denote as $\tilde{\bm{x}}=[\tilde{x}^{(1)},\dots,\tilde{x}^{(d_T-1)}]^T$. We will from here onwards denote the original (noisy) and reconstructed (denoised) state variables as $\tilde{\bm{x}}$ and $\bm{x}$, respectively. The surface of section for the noisy data defines a noisy mapping of the form
\begin{align}
\tilde{\bm{x}}_{n+1}\approx\bm{F}(\tilde{\bm{x}}_n),\label{eq:02}
\end{align}
where $\tilde{\bm{x}}_n$ represents the $n^{\text{th}}$ piercing of the surface of sections by the noisy data and $\bm{F}:\mathbb{R}^{d_T-1}\to\mathbb{R}^{d_T-1}$ is an unknown mapping function which we assume is continuously differentiable and encodes the hypothesized low dimensional dynamics of the system, which in the infinite $N$ limit would follow $\bm{x}_{n+1}=\bm{F}(\bm{x}_n)$. The noise inherent in the data corrupts this mapping, and therefore our goal is to use the dataset to find an approximation $\bm{F}_{\text{loc}}$ to the mapping function $\bm{F}$ to enable us to reconstruct the denoised dynamics via the evolution
\begin{align}
\bm{x}_{n+1}=\bm{F}_{\text{loc}}(\bm{x}_n).\label{eq:03}
\end{align}

Our method for making this forecast using the noisy data is based on the behavior of nearby noisy points. Consider a point $\bm{x}_n$ for which we wish to predict $\bm{x}_{n+1}$. We thus require an accurate estimation of the mapping function $\bm{F}$. At any given iterate, we do not need a {\it global} estimation, but rather a {\it local} estimation. To make such a local estimation we consider those points $\tilde{\bm{x}}_n$ in the noisy data that lie within a small distance $\epsilon$ (using the $\ell^2$ norm) of our current point $\bm{x}_n$, as well as their forward mappings $\tilde{\bm{x}}_{n+1}$. In Fig.~\ref{fig2} we illustrate a collection of these noisy mappings $\tilde{\bm{x}}_n\mapsto\tilde{\bm{x}}_{n+1}$ in the example presented above. Note that all pre-mapped points $\tilde{\bm{x}}_n$ lie within a circle, while the post-mapped points $\tilde{\bm{x}}_{n+1}$ lie roughly within an ellipse that illustrates the stretching and contraction of the mapping. To best estimate the local behavior of $\bm{F}$ near $\bm{x}_n$, we note that any one mapping $\tilde{\bm{x}}_n\mapsto\tilde{\bm{x}}_{n+1}$ is corrupted by noise, but we hypothesize that on average this noise averages out. Thus, we look for the local linear representation of $\bm{F}$ given by
\begin{align}
\bm{F}(\bm{x})\approx\bm{F}_{\text{loc}}(\bm{x})=\bm{F}(\bm{x}_n)+DF(\bm{x}_n)\bm{x},\label{eq:04}
\end{align}
where we compute the entries of the vector $\bm{F}(\bm{x}_n)$ and matrix $DF(\bm{x}_n)$ using least squares with the nearby points $\tilde{\bm{x}}_{n}$ and their forward mappings $\tilde{\bm{x}}_{n+1}$.

\begin{figure}[t]
\centering
\epsfig{file =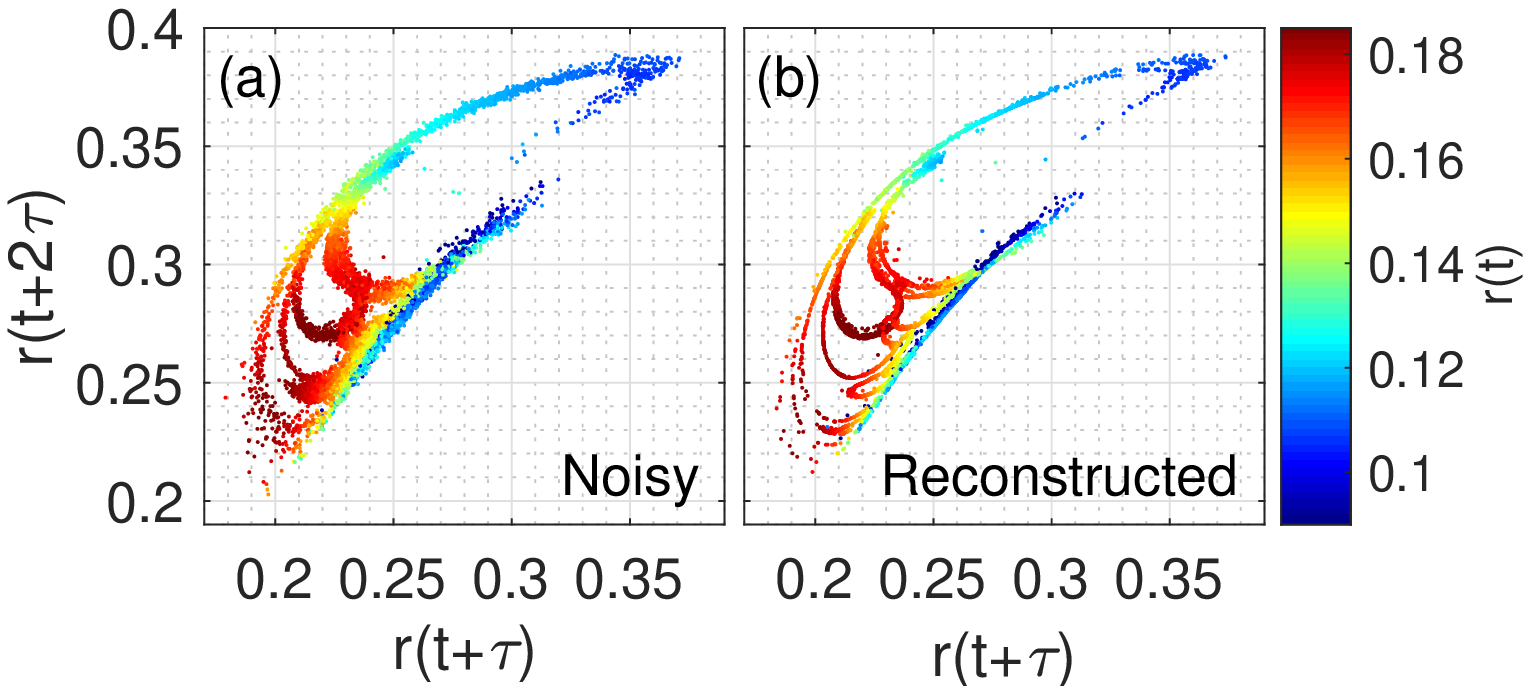, clip =,width=\linewidth }
\epsfig{file =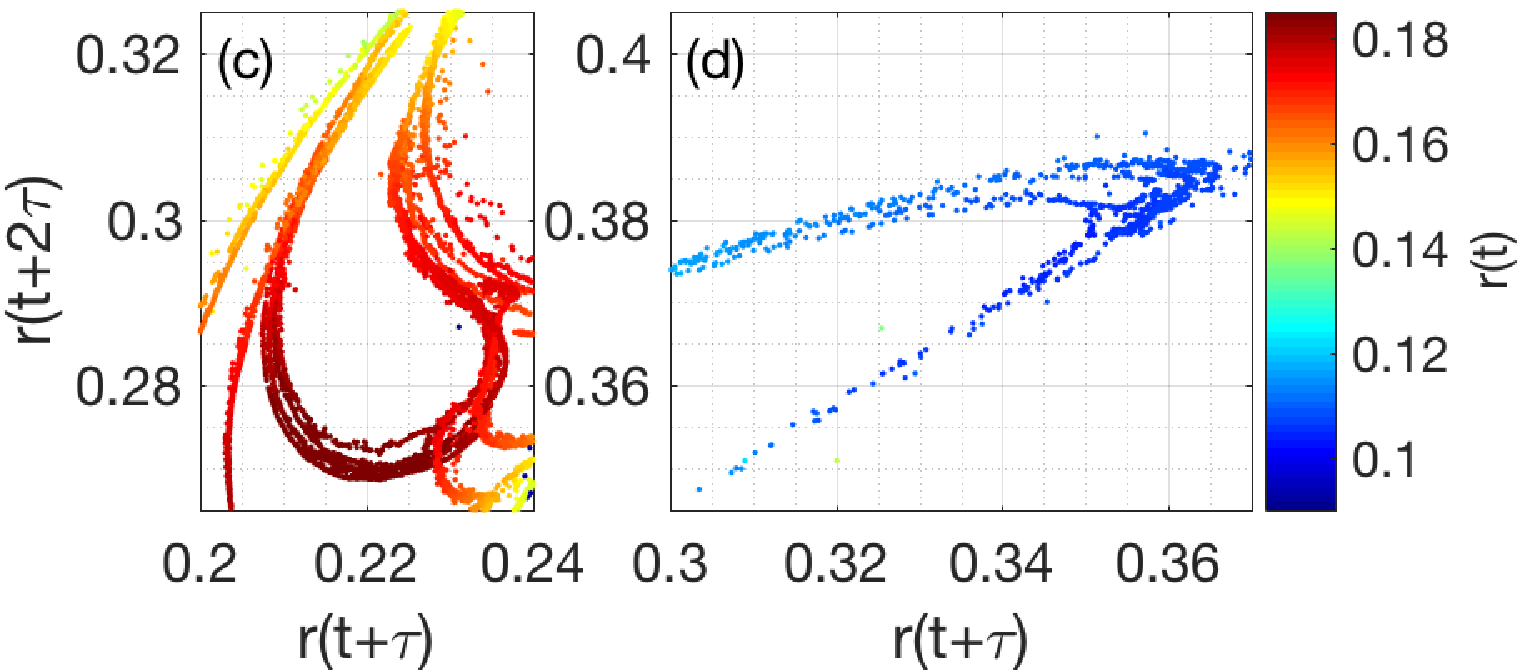, clip =,width=\linewidth }
\caption{(Color online) (a) Noisy and (b) reconstructed dynamics found using a distance threshold of $\epsilon=10^{-3}$ for a system of $N=10^3$ oscillators. (c), (d) Zoomed-in views on the reconstructed dynamics that illustrate greater detail recovered by the reconstruction.}\label{fig3}
\end{figure}

Our algorithm for reconstructing the low dimensional, denoised dynamics is then summarized as follows. Given an initial condition $\bm{x}_0$ we collect each point $\tilde{\bm{x}}_{n}$ from the noisy data that lies within a distance of $\epsilon$ from $\bm{x}_0$, as well as their forward mappings $\tilde{\bm{x}}_{n+1}$. Next, using these collected noisy mappings $\tilde{\bm{x}}_n\mapsto\tilde{\bm{x}}_{n+1}$ we calculate the entries of the vector $\bm{F}(\bm{x}_0)$ and matrix $DF(\bm{x}_0)$ given in Eq.~(\ref{eq:04}) that minimizes $\|\tilde{\bm{x}}_{n+1}-\bm{F}_{\text{loc}}(\tilde{\bm{x}}_n)\|^2$ averaged over the points, where $\|\cdot\|$ is the Euclidean norm. We then determine the image $\bm{x}_1$ of $\bm{x}_0$ using the local description, i.e., $\bm{x}_1=\bm{F}_{\text{loc}}(\bm{x}_0)$. We repeat this process until we collect as many iterates of the reconstructed dynamics as we desire.

We illustrate the utility of the method using our example described above. Specifically, we consider a noisy dataset of $10^6$ points generated by a system of $N=10^3$ oscillators. In Figs.~\ref{fig3}(a) and (b) we plot, for comparison, the original noisy data and the reconstructed dynamics, respectively, in both cases using $10^4$ points. Immediately, we observe a significant reduction in (b) compared to (a). In Figs.~\ref{fig3}(c) and (d) we zoom in on two areas of the reconstructed dynamics to illustrate the added detail picked up in the reconstructed dynamics. 

The implementation of this method requires a few important practical considerations. The primary algorithmic parameter is the distance threshold $\epsilon$ we use to collect noisy points for use in the least squares calculation at each iteration. In principle, $\epsilon$ should be small enough so that the dynamics of all collected points are well approximated by a linear map, but not so small that the number of points used in the least squares calculation is too few. Thus, an appropriate choice of $\epsilon$ is a trade-off between these two constraints and will depend on the size of the dataset available. In the results presented above we used $\epsilon=10^{-3}$, but we note that similar results were obtained for other values of $\epsilon$ (not shown). Moreover, we find that it is helpful to set a minimum on the number of points used in the least squares calculation. That is, if too few points $\tilde{\bf{x}}_n$ are found within a distance $\epsilon$ of $\bm{x}_n$, we include the next closest points to satisfy this lower bound. In our results we require at least $20$ points be used in the calculations. We emphasize that, although we used the local least-squares denoising method, other denoising methods could have been used~\cite{Kantz2003}. The issues regarding choices for $\epsilon$ and the number of neighbors are well known in these methods~\cite{Kantz2003}.

\section{Attractor Invariants}\label{sec4}
We now demonstrate that the reconstruction of the low dimensional dynamics can be used to calculate the dynamical invariants of the attractor in the low dimensional setting. We first consider the Lyapunov spectrum of the strange attractor which for a $d$-dimensional dynamical system consists of the $d$ values $\lambda_1\ge\lambda_2\ge\cdots\ge\lambda_d$ which measure the exponential stretching and contraction of perturbations throughout the attractor and are given by
\begin{align}
\lambda=\lim_{n\to\infty}\frac{1}{n}\ln\frac{|\bm{u}_n|}{|\bm{u}_0|},\label{eq:05}
\end{align}
where $\bm{u}_n$ is one of the tangent vectors evolving according to $\bm{u}_{n+1}=DF(\bm{x}_n)\bm{u}_n$. While our dynamical reconstruction does not yield a full description of the mapping function, the local description given in Eq.~(\ref{eq:04}) suffices to compute the spectrum of Lyapunov exponents for the low dimensional dynamics using typical numerical methods~\cite{Ott2002} since the Jacobian $DF$ is estimated. Using $5\times10^4$ iterations we calculate the full spectrum of eigenvalues for the reconstructed dynamics described above and summarize the values in Table~\ref{table:01}.
 
We next consider the Lyapunov dimension of the attractor, which is defined by the Lyapunov spectrum~\cite{Kaplan1979,Ott2002}. The Lyapunov dimension is defined as
\begin{align}
d_L=k+\frac{\sum_{j=1}^k\lambda_j}{|\lambda_{k+1}|},\label{eq:06}
\end{align}
where $k$ is the largest integer such that the quantity $\sum_{j=1}^k\lambda_j\ge0$. In the case of a system with $\lambda_1>0>\lambda_2$ and $|\lambda_2|>|\lambda_1|$, as ours is, this reduces to $d_L=1+\lambda_1/|\lambda_2|$. Given in Table~\ref{table:01}, this is our first fractal dimension for the strange attractor.

 \begin{table}[t]
 \caption{\label{table:01} Strange attractor invariants computed with the reconstructed dynamics: Lyapunov spectrum and fractal dimensions.}
 \begin{tabular}{ | c | c | c | }
 \hline
 ~Quantity~ & ~Notation~ & ~Value~ \\
 \hline
  \hline
 ~Maximal Lyapunov exponent~ & $\lambda_1$ & $0.2146$ \\
 \hline
 Second Lyapunov exponent & $\lambda_2$ & ~$-0.3882$~ \\
 \hline
  Third Lyapunov exponent & $\lambda_3$ & $-0.9181$ \\
 \hline
 \hline
Lyapunov Dimension & $d_L$ & $1.5528$ \\
 \hline
 Information Dimension & $d_1$ & $1.5675$ \\
 \hline
 Correlation Dimension & $d_2$ & $1.4801$ \\
 \hline
 \end{tabular}
 \end{table}

We next consider two other fractal dimensions, the information and correlation dimensions~\cite{Russell1980PRL,Farmer1983PhysD}. Formally, these dimensions are measured by partitioning the domain of the attractor into $N(\epsilon)$ cubes each of unit size $\epsilon$ and calculating the fraction of time $\mu_i$ spent by typical orbits in each box $i$. The information and correlation dimensions are defined, respectively, by
\begin{align}
d_1=\lim_{\epsilon\to0^+}\frac{\sum_{i=1}^{N(\epsilon)}\mu_i\ln\mu_i}{\ln \epsilon}\text{ and }d_2=\lim_{\epsilon\to0^+}\frac{\sum_{i=1}^{N(\epsilon)}\mu_i^2}{\ln \epsilon}.\label{eq:07}
\end{align}
We proceed by calculating these quantities using the methods outlined in Refs.~\cite{Grassberger1983PRL,Brandstater1987PRA} with $10^5$ points generated by the reconstructed dynamics and report them in Table~\ref{table:01}.

Finally, we comment briefly on the Kaplan-Yorke conjecture~\cite{Kaplan1979}. The Kaplan-Yorke conjecture states that for typical systems (i.e., systems that are not pathologically engineered) the information dimension is equal to the Lyapunov dimension, $d_1=d_L$. In the results obtained for our primary example we find that in fact these two quantities are approximately equal, and since the quantities reported in Table~\ref{table:01} are subject to typical numerical inaccuracies, we judge our results to be in agreement with the Kaplan-Yorke conjecture. Also, as expected, $d_2<d_1$.

\section{Discussion}\label{sec5}

In this paper we have shown the potential of denoising methods for uncovering low dimensional macroscopic dynamics that emerge in large systems of coupled dynamical units. Analytical methods for describing such low dimensional behavior have been found for certain cases of sinusoidally-coupled phase oscillators~\cite{Watanabe1993PRL,Ott2008Chaos}, but in more complicated cases, e.g., most systems of limit cycle oscillators, such analytical methods are not known. Therefore, the identification of such low dimensional behavior, particularly in systems that exhibit rich dynamics and finite-size fluctuations remains an important task. Our method for uncovering low dimensional behavior in the macroscopic system dynamics uses techniques originally designed for denoising chaotic time series to reconstruct the desired low dimensional dynamics. We have used as our primary example a system of globally coupled Landau-Stuart oscillators in the chaotic regime. We have shown that our method not only reconstructs the low dimensional behavior that describe the macroscopic dynamics of the system, but allows for the accurate computation of dynamical invariants, e.g., fractal dimensions and the Lyapunov spectrum.

\acknowledgments

The work of E. O. was supported by ARO Grant W911NF-12-1-0101. J. G. R. acknowledges a useful discussion with Joshua Garland. The authors wish to thank Steven H. Strogatz for an initial conversation that was a motivation for this work.

\bibliographystyle{plain}

\end{document}